# Can a single molecule trap the electron? [1]

Ilya A. Shkrob [1]* and John A. Schlueter [2]

*Chemistry [1] and Materials Science [2] Divisions,*

*Argonne National Laboratory, Argonne, IL 60439*

**Abstract**

We suggest that it might be possible to trap the electron in a cavity of a macrocycle molecule, in the same way this trapping occurs cooperatively, by several solvent molecules, in hydroxylic liquids. Such an encapsulated electron is a 'molecular capacitor,' in which the excess electron is largely decoupled from valence electrons in the trap. A specific design for such a trap that is based on calix[4]cyclohexanol is discussed in detail. It is shown theoretically, by *ab initio* and density functional theory (DFT) modeling, that one of the conformations of this molecule forms the optimum tetrahedral trap for the electron. The resulting 'encapsulated electron' strikingly resembles the solvated electron in alcohols and water.

________________________________________________________________

* To whom correspondence should be addressed: *Tel* 630-252-9516, *FAX* 630-2524993, *e-mail:* shkrob@anl.gov.

# 1. Introduction

In hydroxylic liquids and amorphous solids, such as water (ice) and alcohols, the excess electron (known as 'solvated' or 'trapped' electron) occupies a small cavity formed by 4-8 polar solvent molecules. [1-4] This state is stabilized by attraction to several OH groups pointing towards the center of the cavity and repulsion by hydroxyl protons via the Pauli exclusion. Our recent results [5,6] suggest that this familiar picture of the solvated/trapped electron might be incomplete: in addition to the electron density inside the cavity, there is a substantial (ca. 18% for hydrated electron, $e^-_{hyd}$ [5]) density in the *O 2p* orbitals of the oxygen atoms in the solvating OH groups. Despite this sharing, the excess electron is largely decoupled from valence electrons in the solvent molecules. The resulting structure is perhaps the closest approximation to 'a particle in a box' that one comes across in a chemical system. This particular 'box' is the smallest possible for a molecular system: e.g., in water, the cavity is less than 5 Å in diameter. Due to the relative isolation of the excess electron from the solvent molecules (including the magnetic nuclei in these molecules), solvated electrons exhibit very long times of spin dephasing, even at room temperature (microseconds). In this regard, the solvated electrons appear to be superior to other embodiments of nanoscale 'boxes' (e.g. semiconductor quantum dots) suggested for, *inter alia*, quantum computing which often exhibit short decoherence times, even at low temperature (nanoseconds at 4 K).

Experiments on electron trapping by self-assembled clusters of alcohol molecules in liquid hydrocarbon solvents [7] and alkane glasses suggest that three or four OH groups are already sufficient to bind the electron at ca. 0.8 eV below the conduction band of the solvent; the absorption spectra of electrons trapped by such clusters are also very similar to those observed for solvated electrons in neat alcohols. Importantly, the equilibrium fraction of such tetramer electron traps in the reaction mixture might be small yet such H-bonded clusters are the only deep traps for the electron in these binary solutions. According to electron spin echo envelope modulation (ESEEM) studies of Kevan et al. [8] the mean coordination number for the trapped electron in low temperature ethanol glass is four. Electron nuclear double resonance (ENDOR) spectroscopy of electrons trapped by single crystals of carbohydrates (such as sucrose) indicates that either two OH groups or, more likely, two magnetically equivalent sets of such OH groups stabilize the excess negative charge. [9] The recent *ab initio* and density functional theory (DFT) calculations also indicate that four non-H-bonding hydroxyl groups suffice for trapping the electron in the interior of a water cluster of 20-24 molecules; [10] the remaining water molecules serve as 'dielectric continuum' and provide an H-bond network that fixes the four solvating molecules in the orientation that facilitates electron trapping by dangling OH groups (see below).

These considerations suggests that four OH groups arranged on the tetrahedral pattern would form a sufficiently deep trap for electron binding. In hydroxylic solvents and clusters, such arrangements are the result of fortuitous orientation of fluctuating OH dipoles. However, a suitably designed (semi-)rigid single molecule that has an internal cavity resembling the tetrahedral electron-trapping cavity in these hydroxylic solvents should also be suitable. Such a molecule would trap ('encapsulate') the electron in the same fashion that several molecules cooperatively trap it in polar liquids and crystals. The resulting anion would hold the excess negative charge primarily in its cavity, where the electron wavefunction is minimally perturbed by the electrons in the molecule. Such an anion would be the chemical realization of a 'particle in a box,' thereby constituting the ultimate in the miniaturization of a capacitor. Below we suggest a specific design for such a molecule that is based on perhydrogenated calix[4]arenes.

**2. Calix[2]cyclohexanol.**

The synthesis of *RsS SsR RsS SsR* calix[4]cyclohexanol (**1,** Figure 1a) has been given by Biali and co-workers. [11] This macrocycle is a ring of cyclohexanol molecules connected by methylene linkages. This macrocycle is obtained by hydrogenation of the parent compound, calix[4]arene. In a crystal, the molecule assumes a 'cone' conformation (similar to that of the parent compound) with four OH groups lying in the same plane forming a closed ring (Figure 1a); these OH groups are held in the ring by H-bonds (the O-O distance, 2.71 Å is close to such a distance in $I_h$ ice, 2.77 Å). Such H-bonded rings cannot trap the electron, as suggested by *ab initio* and DFT calculations for tetramers of hydroxylic molecules [12] and our calculations for cone conformer **1.** However, in solution **1** constantly undergoes cone-to-cone inversion through the annulus of the cone, by concerted rotation around four methylene groups. At 100 °C, this ring inversion may be observed directly using NMR, via the exchange of 'inward' and 'outward' pointing methine protons; the inversion barrier is ca. 1 eV. [11] At the outset of the inversion, the H-bound ring of the four OH bonds is broken and two cyclohexanol rings flip below the mid plane, yielding conformer **2** (Figure 1b) in which the four hydroxyl groups are at the corners of a tetrahedron. We believe that this $C_{2v}$ symmetrical conformer will be shown to trap electrons, yielding radical anion **2,**$^-$ that strongly resembles the solvated electron in alcohols and water. The electron is expected to be attached to **2** in the same fashion it can be attached to a cluster of alcohol molecules, [6,7] i.e., by dispersal of the electron trap in a polar (e.g., hexamethylphosphoramid) [13] or nonpolar (alkane) [6] liquid/solid with very low binding energy for the excess electron. The solvated/quasifree electron, that is generated photo- or radio- lytically, in such solvents rapidly descends into the traps. [6,7]

### 3. Computational Details

The electronic structure of **2**$^-$ was studied computationally using a self-consistent field Hartree-Fock (HF) model and a DFT model with B3LYP functional (Becke's exchange functional [14] and the correlation functional of Lee, Yang, and Parr) [15] from Gaussian 98. [16] The basis set was a 6-31G split-valence double-ζ Gaussian set augmented with diffuse and polarized (d,p) functions (6-31+G** or 6-311++G**). In some calculations, this basis set was augmented with additional diffuse functions [10] for the four hydroxyl groups (this set is abbreviated as 6-31(OH1+)+G** in the following).

Isodensity maps for the singly occupied molecular orbital (SOMO) indicate that the electron wavefunction inside the cavity and in the frontier *p*-orbitals of oxygen atoms have opposite signs. Plotting the isodensity contour maps of the SOMO for progressively increasing amplitudes helps to visualize these two contributions (as shown in Figure 2). For **2**[-], the diffuse, positive part of the SOMO occupies 90% of the geometrical cavity at the density of +0.05 $a_0^{-3}$ and 10% at the density of +0.11 $a_0^{-3}$ (where $a_0 \approx 0.53$ Å is the atomic unit of length). The highest (negative) density is found in the frontal *O 2p* orbitals of oxygen atoms of the four hydroxyl groups. The total spin density in the *O 2p* orbitals was defined as explained in ref. [5].

## 3. Results and Discussion

Isodensity maps of the SOMO shown in Figure 2 suggest that most of the electron density is inside the cavity and in the frontal *O 2p* orbitals of the four hydroxyl groups forming the cavity. These SOMO maps may be compared with the maps for geometry optimized $D_{2d}$ symmetrical $(H_2O)_4^-$ and $(MeOH)_4^-$ tetramer anions, shown in Figure 3a and 3b, respectively, that exhibit the same structural motif, and to large water cluster anions (e.g., 24-mer **$5^{12} 6^2$ B** anion of Herbert and co-workers [10]) that trap the electron internally in their central cavities (Figure 3c). At the B3LYP/6-311+G** level, all of these cluster anions have ca. 10% of the spin density in the *O 2p* orbitals, despite the considerable variation in the cavity size $r_{XH}$ (defined as the mean distance to the hydrogens in the solvating OH groups measured from the center of gravity of the SOMO, Table 1). For **2**[-], $r_{XH}$ is only 1.35 Å (B3LYP models) or 1.55 Å (HF models) vs. 1.8-2.2 Å obtained for these model clusters. In mixed quantum/classical molecular dynamics (QC/MD) models of $e_{hyd}^-$, the most probable $r_{XH}$ distances are 2.0-2.2 Å. [3,5] For electrons trapped in alkaline glasses, this distance is 2-2.1 Å, [17,18] in ethanol glass, ca. 2.2 Å [8], in carbohydrate crystals (dulcitol, arabinose, sucrose, etc.) - between 1.59 and 1.82 Å. [9]

Figure 4 shows the angle average radial component of the SOMO density for **2**[-] geometry optimized using the B3LYP/6-31(OH1+)+G** model. The most probable location of the electron is inside the cavity (at $r \approx 1$ Å). This SOMO wavefunction can be approximated by an *s*-type hydrogenic wavefunction $\Psi_s(r) \propto (\pi/\lambda^3)^{1/2} \exp(-r/\lambda)$, with $\lambda \approx 1.15$ Å (vs. 1.7 Å for $e^-_{hyd}$ in the DFT-QC/MD model [5]). The radius of gyration is $r_g \approx 2.4$ Å (defined as $r_g^2 = \langle r^2 \rangle$) compared with 2.74 Å for $e^-_{hyd}$. [5] The semiaxes of the gyration ellipsoid (whose squares are the principal values of the symmetrical $\langle r_\alpha r_\beta \rangle$ tensor, where $\alpha, \beta = x, y, z$) are 1.17 Å x 1.47 Å x 1.47 Å (vs. 1.48 Å x 1.58 Å x 1.69 Å for $e^-_{hyd}$ [5]) i.e., the isodensity surfaces of the electron wavefunction are, to a first approximation, oblate ellipsoids. The cavity (that is, the sphere of radius $r_{XH}$) contains ca. 40% of the unpaired electron density (vs. ca. 60% in $e^-_{hyd}$ [5]), and the negative lobes of the frontal *O 2p* orbitals contain ca. 26% of the density (vs. 12% for $e^-_{hyd}$); the total population of these orbitals is ca. 30% (vs. 18% for $e^-_{hyd}$). Thus, although **2**- has more anionic character than $e^-_{hyd}$ and 30% tighter localization, the electronic properties are similar to those for solvated/trapped electrons in ordinary hydroxylic solvents: the wavefunction is more-or-less hydrogenic, and the cavity is slightly asymmetrical.

A convenient way to characterize spin density distribution in a solvated electron is through the hyperfine coupling constants (hfcc's). These constants, unlike the electron density profiles, can be determined directly using electron paramagnetic resonance (EPR) and ESEEM spectroscopies. As seen from Table 1, cluster anions shown in Figure 3 exhibit large isotropic hfcc on $^{17}$O nuclei in the solvating OH groups; in this regard, **2**- also resembles the solvated electron (e.g., for $e^-_{hyd}$ the hfcc constant is ca. -15 G [5]). The small hfcc's for OH protons in the DFT models are due to the spin bond polarization involving the *O 2p* orbitals: this effect almost cancels the positive contribution from the in-cavity wavefunction (which is in contact with these protons). In HF models, this kind of compensation does not occur, and these models yield large negative proton coupling constants (Table 2). [5] Hfcc's for other magnetic nuclei ($^1$H and $^{13}$C) are small, the largest proton coupling is for methylene protons in *β*–position to OH groups; this coupling is due to hyperconjugation to the *O 2p* orbital. For normal isotopic abundances, the estimated second moment [17] of the EPR spectrum (for DFT models, Table 2) would be ca. 89 G$^2$ (vs. 21-23 G$^2$ for $e^-_{hyd}$ [5,17]); the largest contribution to this parameter is from the four hydroxyl protons due to their large anisotropic hfcc of ≈12.8 G (vs. ≈6 G for the six closest protons in $e^-_{hyd}$ [5,18]). Due to the high symmetry of the trap, the powder EPR spectrum is predicted to exhibit well resolved structure, dominated by a pentet of resonance lines separated by 10 G (Figure 5). By contrast, EPR spectra of solvated/trapped electrons are not resolved, due to the multiplicity of trapping sites and large variation of $^1$H coupling constants within a given site. This characteristic 5-line EPR spectrum might constitute the proof of the suggested structure for **2**$^-$.

The estimate for the gas-phase adiabatic electron affinity $EA_g$ of **2** is 120 meV (in the B3LYP/6-31+G** model) or 110 meV (in the B3LYP/6-31(OH1+)+G** model). In a hydrocarbon solution, the binding energy $EA_{liq}$ of the electron in a trap is given by $EA_{liq} \approx EA_g - P_- + V_0$, where $P_- \approx -e^2(1-1/\varepsilon)/2a$ is the Born polarization energy of the anion (where $e$ is the electron charge, $\varepsilon \approx 2$ is the static dielectric constant, and $a$ is the radius of the anion) and $V_0$ is the energy of the quasifree electron in a liquid vs. vacuum (which is close to zero for *n*-alkanes [19]). For $a \approx 4.35$ Å and $V_0 \approx +100$ meV (*n*-hexane), [19] one obtains $P_- \approx$ -830 meV and $EA_{liq} \approx 1$ eV: that is, **2** is a deep electron trap in the alkane solution.

### 4. Conclusion.

We predict that conformer **2** of calix[4]hexanol would trap electrons in essentially the same way as liquid/solid water/ice and alcohols. The resulting anion species, **2**$^-$, has a highly regular, symmetrical structure that facilitates advanced spectroscopic studies. In hydrocarbon solutions, **2** is a deep electron trap of ca. 1 eV. The electron is well localized inside the tetrahedral cavity, with ca. 30% of the spin density in the *O 2p* orbitals of the four OH groups. The resulting structure may be considered as a nanoscale molecular capacitor holding the unit negative charge. Electron encapsulation by macrocycles examined in this Letter can be viewed as further elaboration of dipole coagulation [20], semicontinuum [21], and snowball [22] models for electron localization in dispersed polar clusters that were developed in the 1970s.

**5. Acknowledgement.**

IAS thanks C. D. Jonah, J. R. Miller, J. F. Wishart, M. Tachiya, and F. T. Williams for many useful discussions and F.-P. Schmidchen, J.-M. Lehn, J. L. Dye, J. Sessler, and E. Anslyn for their advice. This work was performed under the auspices of the Office of Science, Division of Chemical Science, US-DOE under contract number W-31-109-ENG-38.

**Table 1**

Isotropic hyperfine coupling constants and cavity size for **2** and model cluster anions (B3LYP/6-31+G** model).

|  | **2⁻** | $(H_2O)_4^-$ | $(MeOH)_4^-$ | $(H_2O)_{24}^-$ [b] |
|---|---|---|---|---|
| $A(^1H)$, G [a] | 0.32 *(0.25)* | -0.67 | 0.17 (-0.14) | -0.36 *(-0.65)* |
| $-A(^{17}O)$, G | 29.3 *(29)* | 43.9 *(19)* | 25.8 (15.9) | 17.2 *(16.3)* |
| $r_{XH}$, Å | 1.34 (1.36) | 1.57 *(1.96)* | 1.8 *(2.21)* | 2.16 |

a) 1 Gauss = 10⁻⁴ Tesla, b) ***5¹² 6² B*** cluster anion [10]. The values in the parenthesis were obtained using the B3LYP/6-311++G** model or the B3LYP/6-31(OH1+)+G** model (for **2⁻**).

**Table 2.**

Isotropic hyperfine constants for magnetic nuclei in **2**[-] (a comparison between the computational models)

| method | B3LYP | | HF | |
|---|---|---|---|---|
| isotropic hfcc's, G | 6-31+G** | a | 6-31+G** | a |
| $^{17}O$ | -29.3 | -29 | -21.3 | -22.7 |
| $^{1}H_O$ | 0.33 | 0.25 | -12.2 | -12.2 |
| $^{13}C_a$ | 0.17 | 0.15 | -0.11 | -0.15 |
| $^{13}C_b$ | 0.39 | 0.51 | 1.0 | 0.96 |
| $^{13}C_cHOH$ | -0.6 | -0.49 | -0.9 | -0.9 |
| $^{13}C_d$ | 1.3 | 1.3 | 0.71 | 0.70 |
| $^{13}C_e$ | 0.3 | 0.3 | - | - |
| $^{1}H_{CHOH}$ | 1.6 | 1.5 | 0.82 | 0.81 |
| $r_{XH}$, Å | 1.34 | 1.36 | 1.56 | 1.55 |

(a) 6-31(OH1+)+G** basis set

**Figure captions.**

**Figure 1.**

(a) The lowest-energy (cone) conformation (**1**) of calix[4]cyclohexanol. Hydrogen bonds between the four OH groups are shown by dashed lines. (b) Electron-trapping conformation **2**: $X$ denotes the center of the cavity; and the OH groups are pointing towards this common center.

**Figure 2.**

Isodensity contours for singly occupied molecular orbital (SOMO) for **2**$^-$ (B3LYP/6-31(OH1+)+G** model). The dark and light gray contours correspond to negative and positive density, respectively; the isodensity levels are (a) ±0.02 and (b) ±0.04 $a_0^{-3}$.

**Figure 3.**

As figure 2, for (a) $(H_2O)_4^-$, (b) $(MeOH)_4^-$, and (c) $5^{12}\ 6^2\ B$ isomer of $(H_2O)_{24}^-$ (B3LYP/6-311++G** model). The isodensity levels are (a,c) ±0.04 $a_0^{-3}$, (b) ±0.05 $a_0^{-3}$.

**Figure 4.**

Bold solid line (to the left): angle integrated radial density of SOMO for **2**$^-$. The dashed line is a fit to a hydrogenic wavefunction. The vertical dash-dot line corresponds to $r = r_{XH}$.

**Figure 5.**

Simulated powder EPR spectrum (to the left) and its first derivative (to the right), for **2**$^-$ anion (B3LYP/6-31(OH1+)+G** model).

**References:**

[1]     E. J. Hart and M. Anbar, "The Hydrated Electron" (Wiley-Interscience: New York, 1970).

[2]     J. Schnitker, P. J. Rossky, and G. A. Kenney-Wallace, J. Chem. Phys. 85 (1986) 2989; T. H. Murphrey and P. J. Rossky, J. Chem. Phys. 99 (1993) 515.

[3]     B. J. Schwartz and P. J. Rossky, J. Chem. Phys. 101 (1994) 6917

[4]     A. Staib and D. Borgis, J. Chem. Phys. 103 (1995) 2642

[5]     I. A. Shkrob, R. Larsen, and B. J. Schwartz, J. Phys. Chem. B (2006).

[6]     I. A. Shkrob and M. C. Sauer, Jr., J. Phys. Chem. A 110 (2006) 8126.

[7]     I. A. Shkrob and M. C. Sauer, Jr., J. Chem. Phys. 122 (2005) 134503.

[8]     M. Narayana and L. Kevan, J. Chem. Phys. 72 (1980) 2891 and refrences therein.

[9]     A. Lund and S. Schlick, Res. Chem. Intermed. 11 (1989) 37.

[10]    J. M. Herbert and M. Head-Gordon, J. Phys. Chem. A 109 (2005) 5217; Phys. Chem. Chem. Phys. 8 (2006) 68.

[11]    I. Columbus, M. Haj-Zaroubi and S. E. Biali, J. Am. Chem. Soc. 120 (1998) 11806.

[12]    H. M. Lee, S. Lee, and K. S. Kim, J. Chem. Phys. 119 (2003) 187, K. S. Kim, S. Lee, J. Kim, and J. Y. Lee, J. Am. Chem. Soc. 119 (1997) 9329.

[13]   R. Catterall, L. P. Stodulski, and M. C. R. Symons, J. Chem. Society A: 1968, 437.

[14]   A. D. Becke, Phys. Rev. A 38 (1988) 3098

[15]   C. Lee, W. Yang, and R. G. Parr, Phys. Rev. B 37(1988) 785.

[16]   M. J. Frisch et al, Gaussian 98, revision A.1, Gaussian, Inc., Pittsburgh, Pennsylvania, 1998.

[17]    P. A. Narayana, M. K. Bowman, L. Kevan, V. F. Yudanov, and Yu. D. Tsvetkov, J. Chem. Phys. 63 (1975) 3365.

[18]    A. V. Astashkin, S. A. Dikanov, and Yu. D. Tsvetkov, Chem. Phys. Lett. 144 (1988) 258.

[19]    R. A. Holroyd, in "Charged Particle and Photon Interactions with Matter," ed. A. Mozumder and Y. Hatano (New York, 2004), pp. 175.

[20]    A. Mozumder, J. Phys. Chem. 76 (1972) 3824.

[21]    K. Fueki, P. A. Narayana, and L. Kevan, J. Chem. Phys. 64 (1976) 4571.

[22]    M. Tachiya, J. Chem. Phys. 69 (1978) 748.

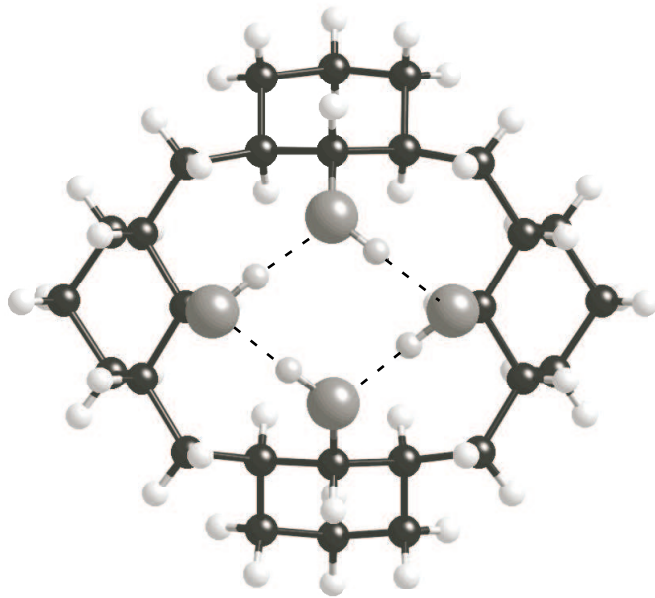

**(a)**

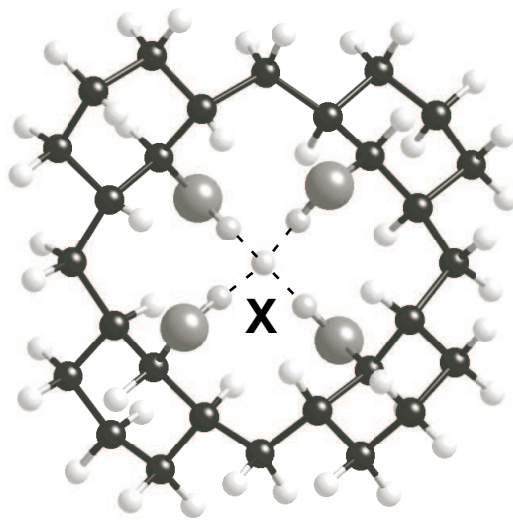

**(b)**

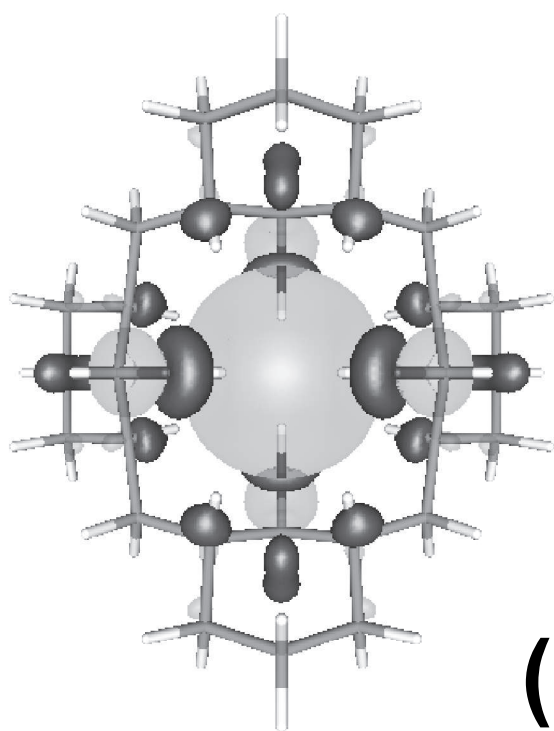
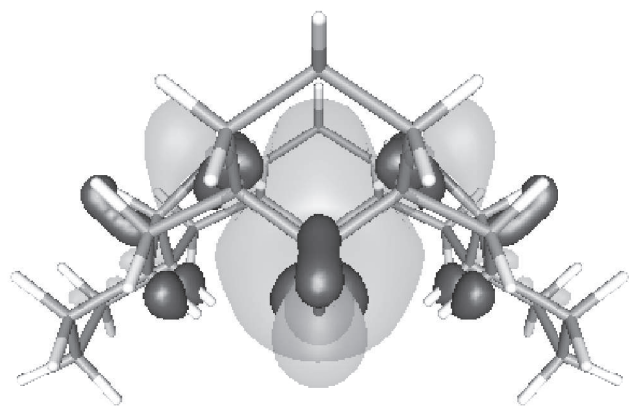

**(a)**

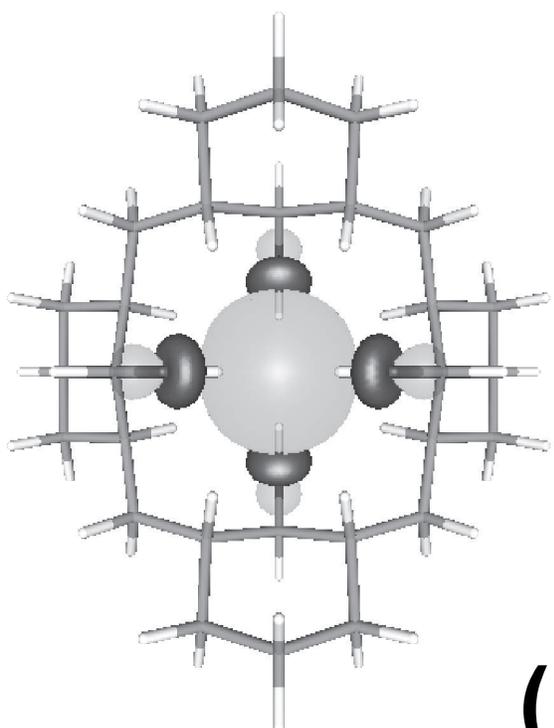
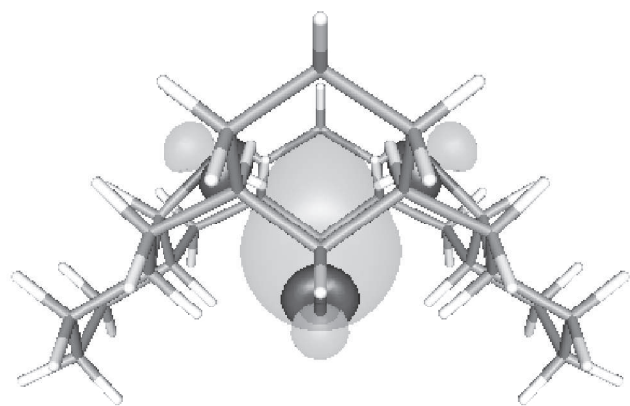

**(b)**

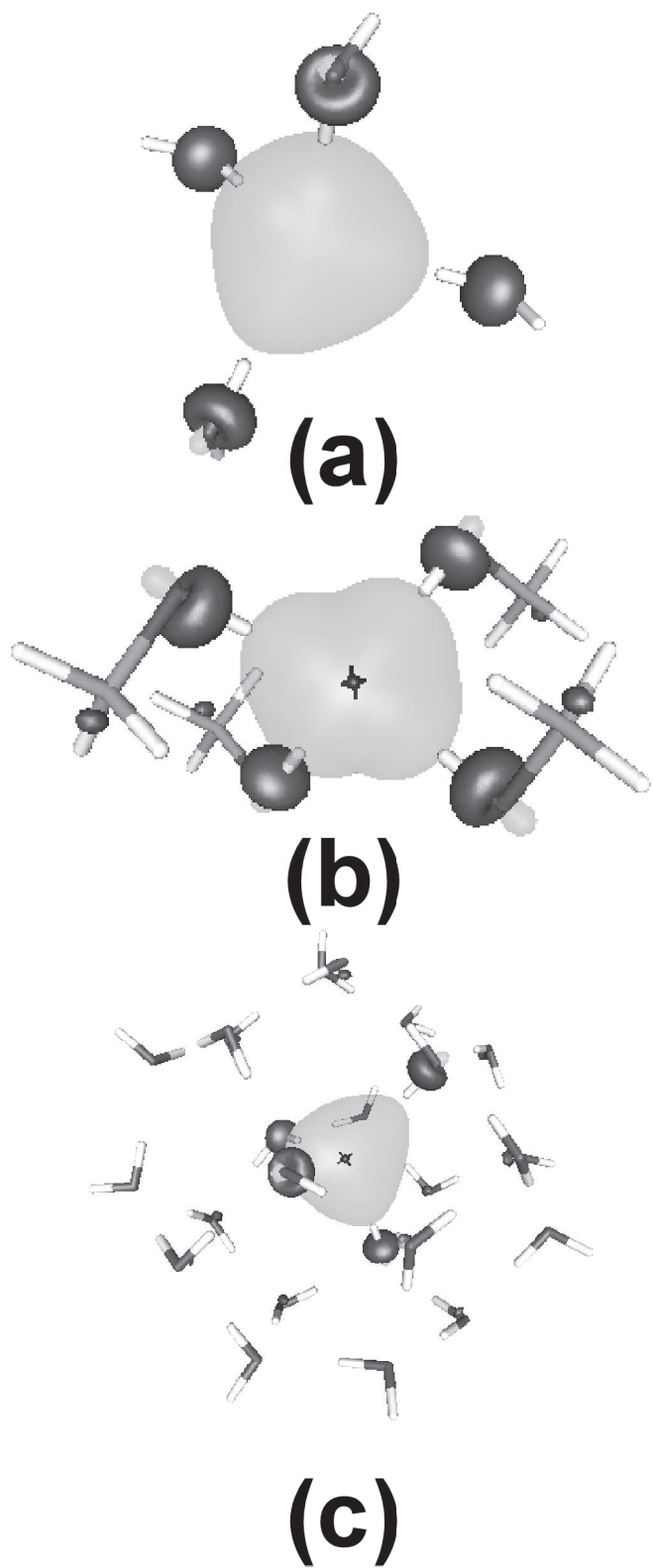

(a)

(b)

(c)

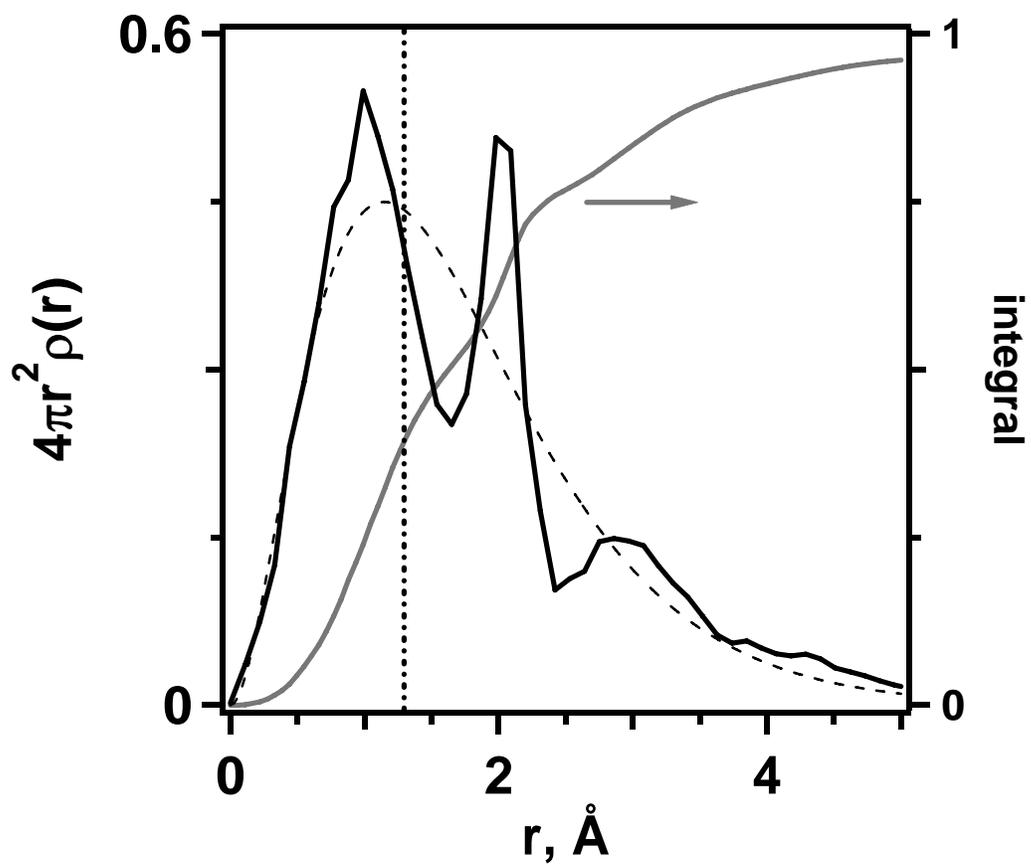

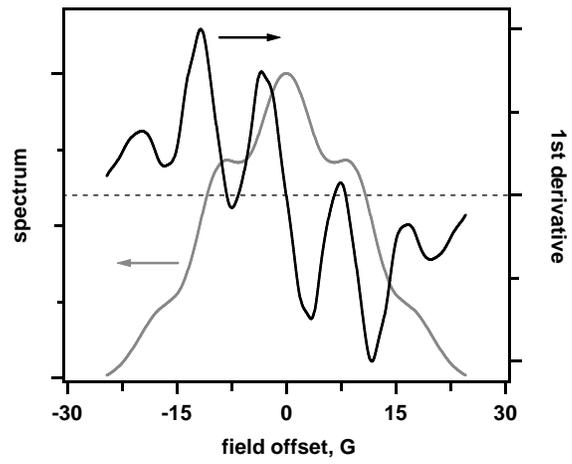